\documentclass[a4paper,11pt]{article}
\usepackage[latin1]{inputenc}
\usepackage[T1]{fontenc}
\usepackage{amssymb}
\usepackage{amsmath}
\usepackage{graphicx}
\usepackage{pstricks}
\usepackage{vmargin}

\setmarginsrb{2.5cm}{2cm}{2.5cm}{2cm}{0cm}{1cm}{0cm}{1cm}

\numberwithin{equation}{section}

\newcommand{\cir}{\!\circ\!} 
\newcommand{\RR}{\mathbb{R}}

\newcommand{\ZZ}{\mathbb{Z}}

\newcommand{\be}{\begin{equation}}
\newcommand{\ee}{\end{equation}}
\newcommand{\bea}{\begin{eqnarray}}
\newcommand{\eea}{\end{eqnarray}}
\newcommand{\ud}{\mathrm{d}}
\newcommand{\G}{\left}
\newcommand{\D}{\right}
\newcommand{\p}{\partial}

\newcommand{\cV}{\mathcal{V}}
\newcommand{\cG}{\mathcal{G}}
\newcommand{\cS}{\mathcal{S}}

\newcommand{\w}{\wedge}

\newcommand{\cY}{\mathcal{Y}}
\newcommand{\cX}{\mathcal{X}}

\newcommand{\cW}{\mathcal{W}}
\newcommand{\cK}{\mathcal{K}}

\begin{document}

\begin{flushright}
ULB-TH/04-34\\
December 2004\\
\vspace*{1cm}
\end{flushright}

\begin{center}
\begin{Large}
\textbf{Selfduality of $d=2$ Reduction\\of Gravity Coupled to a $\sigma$-Model}
\end{Large}

\vspace{7mm}
{\bf Louis Paulot}

\vspace{5mm}
Physique théorique et mathématique, Université libre de Bruxelles\\
and\\
International Solvay Institutes\\
Campus Plaine C.P.~231, B--1050 Bruxelles, Belgium
\vspace{3mm}

{\ttfamily lpaulot@ulb.ac.be}
\end{center}

\vspace{3mm}
\hrule
\begin{abstract} 

Dimensional reduction in two dimensions of gravity in higher
dimension, or more generally of $d=3$ gravity coupled to a
$\sigma$-model on a symmetric space, is known to possess an infinite
number of symmetries. We show that such a bidimensional model can be
embedded in a covariant way into a $\sigma$-model on an infinite
symmetric space, built on the semidirect product of an affine group by
the Witt group. The finite theory is the solution of a covariant
selfduality constraint on the infinite model. It has therefore
the symmetries of the infinite
symmetric space. (We give explicit transformations of the gauge
algebra.) The usual physical fields are recovered in a triangular
gauge, in which the equations take the form of the usual linear
systems which exhibit the integrable structure of the
models. Moreover, we derive the constraint equation for the conformal
factor, which is associated to the central term of the affine group
involved.

\end{abstract}
\hrule

\vspace{3mm}

\section{Introduction}

Dimensional reduction of gravity in dimension $d=4$ to $d=2$ enlarges the
group of symmetries to an infinite group \cite{geroch}, which has been
related to the integrable structure of the theory \cite{maison,bz}. The
symmetry group appeared to be an affine Kac-Moody group \cite{j1,bm1,bm2}.
Furthermore, it was shown that this structure was shared by a full class of
theories, such as supergravities, which in dimension three
reduce (for the bosonic sector) to a symmetric space $\sigma$-model coupled 
to gravity \cite{j1,bm1,n1}.

In addition to the scalars which describe a
$\mathfrak{G}/\mathfrak{H}$ $\sigma$-model, the (bosonic) degrees of freedom
of these models consist of a dilaton $\rho$ and the conformal factor
$\lambda = e^\sigma$ of the metric. The equations of motion
are, in conformal gauge,
\bea
\ud\!*\!\ud \rho &=& 0 \\
\nabla(\rho *\! P ) &=& 0
\eea
($P$ and $\nabla$ are precisely defined below.)
with in addition a first order constraint on the conformal factor,
\be
\p_\pm \rho \, \p_\pm \hat{\sigma} = \frac{1}{2} \rho \, \langle  P_\pm  ,
P_\pm \rangle
\ee
where $\hat{\sigma} = \sigma - \frac{1}{2} \ln\G( \p_+ \rho \, \p_- \rho
\D)$. We choose a Lorentzian spacetime for exposition; it is not
hard to adapt the results for a Euclidean one.

In \cite{ju-ni,be-ju}, the symmetry of such a theory was
enlarged to the full semidirect product of the Witt group and an affine
Kac-Moody group. The fields are infinitely dualised and live in an infinite 
coset
\be
\mathcal{M} = \frac{\cW \ltimes \mathfrak{G}^\infty}{ \cK \ltimes
\mathfrak{H}^\infty} \rlap{\ .}
\ee
In this formalism, the equations of motion come from linear systems which
are imposed as constraints in a triangular gauge.

Here, following \cite{p}, which deals with the flat space $\sigma$-model, 
we restore the infinite $\cK \ltimes
\mathfrak{H}^\infty$ gauge-invariance: we define the finite, constrained
model through a covariant selfduality equation on the infinite tower of
fields. Previously known Lax pairs are recovered as consequences of this
constraint when we go into the triangular gauge. This gives a formulation of
the $d=2$ theory very analogous to the oxidised versions ($d \geq 3$)
\cite{cjlp2,hjp1,dhjp}.

We also derive the constraint for the conformal factor from the selfduality
constraint. Whereas for other fields the duality involves a Hodge
dualisation, it is worth noticing that this is not the case for the
conformal factor, associated to the central term of the group.

In section \ref{section-structure}, we decribe the infinite symmetric space
$\mathcal{M}$ and the algebraic structures involved. In section
\ref{section-selfduality}, we define the duality operator and we show that a
a selfduality constraint can be imposed and is covariant with respect to the
infinite gauge algebra transformations. Finally, we fix the gauge in section
\ref{section-triangular} to recover the physical content of the theory, and
we derive the Lax pair equations associated to the dynamical fields of the
model, together with the conformal factor constraint.

\section{Infinite $\sigma$-model structure}
\label{section-structure}

Following \cite{ju-ni}, 
we consider fields in an infinite-dimensional symmetric space
\be
\mathcal{M} = \frac{\cW \ltimes \mathfrak{G}^\infty}{ \cK \ltimes
\mathfrak{H}^\infty} \rlap{\ .}
\ee
$\cW$ is the ``group'' of diffeomorphisms of the real line, 
$\mathfrak{G}^\infty$ is the affine extension $\mathfrak{G}^{(1)}$ of
the simple group $\mathfrak{G}$ and 
$\cK \ltimes \mathfrak{H}^\infty$ is the subgroup of fixed points of 
$\cW \ltimes \mathfrak{G}^\infty$ under some involution.

Explicitely, $\mathfrak{G}^\infty$ is the set of pairs $(g(t),a)$ where
$g(t)$ is a map from $\RR_+^\times$ to $\mathfrak{G}$
and $a$ is a positive real number. The group law is
\be
(g_1(t),a_1)(g_2(t),a_2) = \G(g_1(t) g_2(t),a_1 a_2 e^{\Omega(g_1,g_2)}\D)
\ee
where $\Omega$ is a group 2-cocycle (see \cite{bm2}).
The Lie algebra is the affine Kac-Moody algebra $\mathfrak{g}^\infty$ 
with analytic functions $b(t)$ with values in $\mathfrak{g}$ and in 
addition a central charge $c$; commutation relations are
\be
\G[b_1(t) , b_2(t) \D] = \G[b_1(t),b_2(t)\D]_\mathfrak{g} 
+ \omega(b_1,b_2) \,c 
\ee
where $\omega$ is a 2-cocycle of the loop algebra (see
\cite{bm2,kac}). Here, it is defined as
\be
\omega(b_1,b_2) = \frac{1}{2}
\oint_{\mathcal{C}_1} \!\!\ud t \, \langle \p_t b_1(t) , b_2(t) \rangle
+ \frac{1}{2} \oint_{\mathcal{C}_2} \!\!\ud t \,
\langle \p_t b_1(t) , b_2(t) \rangle
\ee
where $\mathcal{C}_1$ and $\mathcal{C}_2$ are two contours exchanged
and reversed under $t \rightarrow \frac{1}{t}$, avoiding
singularities \cite{bm2}.

$\cW$ is, at least formally, the group $\mathrm{Diff}^+(\RR_+^\times)$ 
of analytic diffeomorphisms of the real line preserving the orientation. 
We see it as
Laurent series $f(t)$ with the law group given by composition:
\be
(f_1 \cir f_2)(t) = f_1(f_2(t))
\rlap{\ .}
\ee
Its Lie algebra is the real Witt algebra with generators $L_n=t^{n+1}\p_t$ 
($n \in \ZZ$) and commutation relations
\be
\G[ L_m,L_n\D] = (n-m)L_{m+n}
\rlap{\ .}
\ee

The semidirect product $\cW \ltimes \mathfrak{G}^\infty$ 
is given by triples $(f,g,a) \in \cW \times \mathfrak{G}^\infty$
with product law
\be
(f_1,g_1,a_1)(f_2,g_2,a_2) =
 \G( f_1 \cir f_2 \, , \, (g_1 \cir f_2) \, g_2
\, , \, a_1 a_2 e^{\Omega(g_1 \cir f_2 , g_2)} \D)
\rlap{\ .}
\ee

The subgroup $\cK \ltimes \mathfrak{H}^\infty$ consists of fixed
points under an involution $\tau_\ltimes$, which is given by an
involution on $\cW$ 
\be
\tau_\cW : \ f(t) \ \longrightarrow \ \frac{1}{f(1/t)}
\ee
compatible with an involution on $\mathfrak{G}^\infty$
\be
\tau_\infty : \ (g(t),a) \ \longrightarrow \ 
\G( \tau\!\G(g\!\G(\frac{1}{t}\D)\D) , \frac{1}{a} \D)
\rlap{\ ,}
\ee
where $\tau$ is the involution fixing $\mathfrak{H}$. 
Denoting by $\cK$ and $\mathfrak{H}^\infty$ the sets of
fixed points under respectively $\tau_\cW$ and $\tau_\infty$, the
fixed points of $\cW \ltimes \mathfrak{G}^\infty$ under $\tau_\ltimes$
is $\cK \ltimes \mathfrak{H}^\infty$.
It consists of elements $(f,g,1)$ with $f\!\G(
\frac{1}{t}\D) = \frac{1}{f(t)}$ and $g\!\G(\frac{1}{t}\D) = \tau(g(t))$.

On the Lie algebra side, this gives an involution acting on generators as
\be
\tau_\ltimes : \ 
\G\{
\begin{array}{rcl}
L_n & \longrightarrow & -L_{-n} \\
t^n T & \longrightarrow & t^{-n} \tau(T) \\
c & \longrightarrow & -c
\end{array}
\D.
\rlap{\ .} 
\ee

From a field in two dimensions with values in the infinite
dimensional coset,
\be
\cV(x) \in \frac{\cW \ltimes \mathfrak{G}^\infty}{ \cK \ltimes
\mathfrak{H}^\infty}
\rlap{\ ,}
\ee
we derive the pull-back of the Maurer-Cartan form
\be
\cG = \ud \cV \, \cV^{-1}
\label{defG}
\ee
with $\Omega'$ the mixed cocycle defined in \cite{bm2}.
$\cG$ satisfies the Maurer-Cartan equation
\be
\ud \cG = \cG \w \cG
\rlap{\ .}
\ee

\section{Selfduality}
\label{section-selfduality}

We decompose $\cG$ as
\be
\cG = \cX + \cY
\ee
where $\cX$ and $\cY$ are respectively invariant and anti-invariant
under $\tau_\ltimes$. Under a gauge transformation, $\cX$ behaves as a gauge
field, whereas $\cY$ is covariant (see \cite{p} for details). Note that
$\cX$ and $\cY$ are formal series in $t$ and $t^{-1}$ and do not
necessarily make sense as analytic functions of $t$.

In order to impose a selfduality constraint, we define an involution $\cS$ 
acting as
\be
\cS : \ 
\G\{
\begin{array}{rcl}
\alpha \, L_n & \longrightarrow & - * \! \alpha \, L_{1-n} \\
\beta \, t^nT & \longrightarrow & *  \beta \, t^{1-n} \tau(T) \\
\gamma \, c & \longrightarrow & - \, \gamma \, c
\end{array}
\D.
\ee
where $\alpha, \beta, \gamma$ are 1-forms and $T$ is any generator 
of $\mathfrak{g}$.

Our claim is the following: the selfduality constraint
\be
\cS \cY = \cY
\label{self}
\ee
reduces the infinite-dimensional $\sigma$-model to the $d=2$ model obtained
by dimensional reduction of a $d=3$ $\mathfrak{G}/\mathfrak{H}$ $\sigma$-model 
coupled to gravity. Moreover, this constraint is invariant under a global
action of $\mathcal{W} \ltimes \mathfrak{G}^\infty$ on the right and a local
$\mathcal{K} \ltimes \mathfrak{H}^\infty$ gauge transformation on the left. 

The global invariance is trivial: $\cG = \ud \cV \, \cV^{-1}$ 
is invariant under right
multiplication of $\cV$ by any constant group element $\Lambda$
\be
\cV(x) \ \longrightarrow \ \cV(x) \Lambda \rlap{\ .}
\ee

Let us check the gauge invariance, with respect to left multiplication by
elements of $\mathcal{K} \ltimes \mathfrak{H}^\infty$:
\be
\cV(x) \ \longrightarrow \ H(x) \cV(x) \rlap{\ .}
\ee

If we write $\cG$ as
\be
\cG = \sum_{n \in \ZZ} A_n L_n + \sum_{n \in \ZZ} B_n t^n + C \, c 
\label{Gexp}
\ee
with $A_n \in \RR$, $B_n \in \mathfrak{g}$ and $C \in \RR$, we have
explicitely
\be
\cY =
\frac{1}{2}\sum_{n \in \ZZ} (A_n + A_{-n}) L_n 
+ \frac{1}{2}\sum_{n \in \ZZ} \G(B_n - \tau(B_{-n})\D) t^n + C \, c 
\rlap{\ .}
\ee
Solutions of (\ref{self}) are given by
\bea
A_n + A_{-n} &=& 2 *^n \! A_0 \nonumber\\
B_n - \tau(B_{-n}) &=& 2 *^n \! P
\label{se} \\
C &=& 0 \nonumber
\eea
where we decompose $B_0 = P+Q$ with $P$ and $Q$ respectively anti-invariant
and invariant under the involution $\tau$, and with $*^2=1$
in Lorentzian signature.

We consider first an infinitesimal gauge transformation in
$\mathrm{Lie}(\mathcal{K})$: $\delta k = \delta a(x) (L_p - L_{-p})$.
It acts on fields as
\bea
A_n & \longrightarrow & A_n + (n-2p) \, \delta a \, A_{n-p} - (n+2p) \,
\delta a \, A_{n+p} + (\delta_{n,p} - \delta_{n,-p}) \, \ud \delta a \nonumber \\
B_n  & \longrightarrow & B_n + (n-p) \, \delta a \, B_{n-p} - (n+p) \, \delta a
\, B_{n+p} \\
C & \longrightarrow & C \nonumber 
\eea
This leave (\ref{se}) invariant, with
\bea
A_0  & \longrightarrow & A_0 - 4 p \, \delta a  *^p \! A_0 \nonumber \\
P & \longrightarrow & P - 2 p \, \delta a  *^p \! P \rlap{\ .}
\eea

For an infinitesimal transformation $\delta h = \delta b(x) t^p + \tau(\delta
b(x)) t^{-p} \in \mathfrak{h}^\infty$, the fields transforms as
\bea
A_n & \longrightarrow & A_n \nonumber \\
B_n  & \longrightarrow & B_n +  \G[\delta b , B_{n-p} \D] + \G[ \tau(\delta
b) , B_{n+p} \D] + (\delta_{n,p} - \delta_{n,-p}) \, \ud \delta b 
\label{dh} \\
C & \longrightarrow & C + \omega\G( \delta h , B \D) \nonumber
\eea
with $\displaystyle B = \sum_{n \in \ZZ} B_n t^n$.

It is not hard to check that the $A_n$ and $B_n$ parts of (\ref{se}) 
are invariant under these transformations, with
\bea
A_0  & \longrightarrow & A_0 \nonumber \\
P & \longrightarrow & P + \G[ \delta b + \tau(\delta b) \, , *^p \! P \D] 
\rlap{\ .}
\eea

For the central extension, it is more subtle. Using (\ref{se}), we can write
down two expansions of $B$:
\bea
B &=& Q + P + \sum_{n \geq 1} t^n *^n \! P + \sum_{n \geq 1} \G( B_{-n}
t^{-n} + \tau( B_{-n}) t^n \D)  \label{b1} \\
B &=& Q + P + \sum_{n \geq 1} t^{-n} *^n \! P + \sum_{n \geq 1} \G( B_n
t^n + \tau(B_n) t^{-n} \D) \label{b2}
\eea
In order to have some analytic quantity, we require that the last term of
either (\ref{b1}) or (\ref{b2}) can be summed up for some value of $|t| \neq
1$ (it depends on the gauge).
From the structure of this term, it will then be analytic in a annulus
$\theta < |t| < \frac{1}{\theta}$. Let us suppose this is the
case for (\ref{b1}). (The reasonning is analogous in the other case.) For
$\theta < |t| < 1$, the full expansion makes sense as an analytic function, and we 
have
\be
B = Q + \frac{1+t^2}{1-t^2} P + \frac{2t}{1-t^2} *\!P + \sum_{n \geq 1} \G(
B_{-n} t^{-n} + \tau( B_{-n}) t^n \D) \rlap{\ .}
\label{b}
\ee
This is the analytic function which is used to compute the central extension
in (\ref{dh}).

As we have seen, the contour integral used for computing
$\omega(\,\cdot\, ,\,\cdot\,)$ is in fact the average over two contours 
(with the same
orientation) exchanged and reversed under $t \rightarrow \frac{1}{t}$.
The last term of (\ref{b}) is regular in $|t|=1$, so we can take the single
unit circle. For the terms singular in $t = \pm 1$, we choose a pair
of contours
avoiding the singular points; practically, we average on the residue at
$0$ and $\infty$. We find
\be
\omega(\delta h , B) = 0
\ee
and therefore (\ref{se}) is completely invariant under $\delta h$.

The selfduality constraint is thus invariant under the full gauge algebra.

\section{Physical content}
\label{section-triangular}

In order to recover the physical content of the selfdual theory, we fix 
(partially) the gauge for $\cV$ in the following way:
\be
\cV = (\mathrm{id},1,a)(\mathrm{id},g,1)(f,1,1) = (f,g\cir f,a)
\ee
with $f$ and $g$ regular in $t=0$, and $f(0)=0$ (see \cite{ju-ni}).  
This triangular gauge
will allow to recover physical fields of the constrained model as $t=0$
values of fields.

From (\ref{defG}), we get
\be
\cG = \G( \ud f \cir f^{-1} \, , \, \ud g \, g^{-1} + \p_t g \, g^{-1} 
\ud f \cir f^{-1} \, , \,
 \ud a \, a^{-1} - \Omega'(g^{-1}, \ud g \, g^{-1} + \p_t g \, g^{-1}
\ud f \cir f^{-1}) \D) \rlap{\ .}
\label{Gphys}
\ee

As $f$ and $g$ are regular in $0$, we have $A_n =0$ and $B_n=0$ in
(\ref{Gexp}) for $n < 0$.

Defining
\bea
\rho &=& \p_t f \!\mid_{t=0} \nonumber \\
g_0 &=& g \!\mid_{t=0} \\
\hat{\sigma} &=& \ln(a) \nonumber
\eea
we have from (\ref{Gphys})
\bea
A_0 &=& \ud \rho \, \rho^{-1} \nonumber\\
B_0 &=& \ud g_0 \, g_0^{-1} \\
C &=& \ud a \, a^{-1} - \Omega'(g^{-1}, B) \rlap{\ .}\nonumber
\eea

Plugging the solutions (\ref{se}) of the selfduality constraint $\cS \cY =
\cY$ into (\ref{Gphys}), we get
\bea
\ud f \cir f^{-1} &=& \frac{1+t^2}{1-t^2} \, \ud \rho \, \rho^{-1} \, t \p_t +
\frac{2t}{1-t^2} *\!\ud \rho \, \rho^{-1} \, t \p_t
\label{laxf}\\
(\ud (g \cir f) \cir f^{-1}) g^{-1} &=& Q + \frac{1+t^2}{1-t^2} \, P +
\frac{2t}{1-t^2} *\! P 
\label{laxg}\\
\ud \hat{\sigma} &=& \Omega'\G(g^{-1} \, , \, Q + \frac{1+t^2}{1-t^2} \, P +
\frac{2t}{1-t^2} *\! P \D) \rlap{\ .}
\label{dsigma}
\eea
(\ref{laxf}) and (\ref{laxg}) are the linear systems associated to the
$d=2$ reduction of the $d=3$ $\mathfrak{G}/\mathfrak{H}$
$\sigma$-model coupled to gravity \cite{bm1,bm2,ju-ni}. $\rho$ is the
dilaton, and $\sigma = \hat{\sigma} + \frac{1}{2} \ln\G( \p_+ \rho \,
\p_- \rho \D)$ is the Liouville variable. $t=f(s)$ is often called the
``variable spectral parameter''.

The equations of motion of $\rho$ and $g_0$ come from the Maurer-Cartan
equation $\ud \cG = \cG \w \cG$:
\bea
\ud\!*\!\ud \rho &=& 0 \\
\nabla(\rho *\! P ) &=& 0
\eea
with the $\mathfrak{H}$-covariant derivative
$\nabla = \ud + \G[ Q , \ \cdot \ \D]$.

Finally, let us show that (\ref{dsigma}) is the constraint equation for the
conformal factor. First, we turn to conformal coordinates $x^\pm$. Acting on
the right with some constant element, we can always manage to have $g$
infinitesimal in the vicinity of some spacetime point $x_0$: $g = 1 + \delta
g$. In this region,
(\ref{dsigma}) can be written as
\be
\p_\pm \hat{\sigma} = \frac{1}{2} \, \omega\G(- \delta g \, , \, 
\frac{1 \mp t}{1 \pm t} \, P_\pm \D) \rlap{\ .}
\ee
According to the definition of $\omega$, we must take the average of 
the residues of $-\p_t \delta g \, \frac{1 \mp t}{1 \pm t} \, P_\pm $ at $0$
and $\infty$. In this case, it means we must take the residue at $0$ plus
one half of the residue at $\pm 1$ ($\delta g$ is a regular function of $t$
\cite{bm2}). We get
\be
\p_\pm \hat{\sigma} = \frac{1}{2} \, \langle \mp \, \p_t \delta g 
\!\mid_{t=\mp 1} \, , \, P_\pm \rangle \rlap{\ .}
\label{dsigma2}
\ee

From the regularity of $g$ in $t$, we also know that the poles in
(\ref{laxg}) comes from $\p_t g \, g^{-1} \ud f \cir f^{-1}$ on the left
handside, here with $g=1+\delta g$. Using also (\ref{laxf}),
we have
\be
\mp \, \p_t \delta g \!\mid_{t=\mp 1} \p_\pm \rho \, \rho^{-1} = P_\pm 
\rlap{\ .} 
\ee
Combining this with (\ref{dsigma2}), we recover the constraint for the
conformal factor:
\be
\p_\pm \rho \, \p_\pm \hat{\sigma} = \frac{1}{2} \rho \, \langle  P_\pm  ,   
P_\pm \rangle \rlap{\ .} 
\ee

\subsection*{Aknowledgments}

This work is partially supported by IISN - Belgium
(convention 4.4505.86), by the ``Interuniversity Attraction Poles
Programme -- Belgian Science Policy'' and by the European
Commission  FP6 programme MRTN-CT-2004-005104, in which we are
associated to V.U.Brussel.


\begin{thebibliography}{99}

\bibitem{geroch}
{\sc R.~Geroch},
{\sl A Method for Generating New Solutions of Einstein's Equation. 2},
J.\ Math.\ Phys.\  {\bf 13} (1972) 394.

\bibitem{maison}
{\sc D.~Maison},
{\sl Are the Stationary, Axially Symmetric Einstein Equations Completely
Integrable?},
Phys.\ Rev.\ Lett.\  {\bf 41} (1978) 521.

\bibitem{bz}
{\sc V.~A.~Belinsky, V.~E.~Zakharov},
{\sl Integration of the Einstein Equations by the Inverse Scattering Problem
Technique and the Calculation of the Exact Soliton Solutions},
Sov.\ Phys.\ JETP {\bf 48} (1978) 985
[Zh.\ Eksp.\ Teor.\ Fiz.\  {\bf 75} (1978) 1953].

\bibitem{j1}
{\sc B.~Julia}, {\sl Infinite Lie Algebras in Physics}, in
``Johns Hopkins Workshop on Current Problems in Particle Theory'', Baltimore,
1981.

\bibitem{bm1}
{\sc P.~Breitenlohner, D.~Maison}, {\sl Explicit and
Hidden Symmetries of Dimensionally Reduced (Super)Gravity Theories}, in
``Retzbach 1983, Proceedings, Solutions of Einstein's Equations: Techniques
and Result'', eds. C.~Hoensealers and W.~Dietz, Springer (1984) 276
MPI-PAE/PTh 1/84.

\bibitem{bm2}
{\sc P.~Breitenlohner, D.~Maison},
{\sl On the Geroch Group},
Ann. Inst. Poincaré {\bf 46} (1987) 215.

\bibitem{n1}
{\sc H.~Nicolai}, {\sl Two-dimensional Gravities and
Supergravities as Integrable System},  in ``Schladming 1991, Proceedings,
Recent Aspects of Quantum Fields'', eds. H.~Mitter and H.~Gausterer,
Springer (1991) 231.

\bibitem{ju-ni}
{\sc B.~Julia, H.~Nicolai}, {\sl Conformal Internal Symmetry of 2d
$\sigma$-models Coupled to Gravity and a Dilaton}, Nucl. Phys. B {\bf 482}
(1996) 431 [arXiv:hep-th/9608082].

\bibitem{be-ju} {\sc D.~Bernard, B.~Julia},
{\sl Twisted Self-Duality of Dimensionally Reduced Gravity and Vertex
Operators}, Nucl. Phys. B {\bf 547} (1999) 427 [arXiv:hep-th/9712254].

\bibitem{p}
{\sc L.~Paulot},
{\sl Superconformal Selfdual $\sigma$-Models},
JHEP {\bf 0410} (2004) 071
[arXiv:hep-th/0404027].

\bibitem{cjlp2} {\sc E.~Cremmer, B.~Julia, H.~Lü, C.~N.~Pope},
{\sl Dualisation of Dualities II: Twisted Self-duality of Doubled
Fields and Superdualities}, Nucl. Phys. {\bf B535} (1998) 242
[arXiv:hep-th/9806106].

\bibitem{hjp1}
{\sc P.~Henry-Labordère, B.~Julia, L.~Paulot},
{\sl Borcherds Symmetries in M-Theory},
JHEP {\bf 0204} (2002) 049
[arXiv:hep-th/0203070].

\bibitem{dhjp} {\sc Y.~Dolivet, P.~Henry-Labordère, B.~Julia, L.~Paulot},
{\sl Superalgebras of Oxidation Chains}, ULB-TH/04-09, to appear.

\bibitem{kac}
{\sc V.~Kac}, {\sl Infinite Dimensional Lie Algebras}, Cambridge University
Press (1990).

\end{thebibliography}
\end{document}